\documentclass[aip,apl,reprint]{revtex4-2}

\usepackage{graphicx}
\usepackage{stmaryrd}
\usepackage{rotating}
\usepackage{amsmath}
\usepackage{amsfonts}
\usepackage{amssymb}
\usepackage{wasysym}
\usepackage[countmax]{subfloat}
\usepackage{dcolumn} 
\usepackage{bm} 
\usepackage{color}

\usepackage{float}
\usepackage{latexsym}

\begin{document}

\title{Quantum geometry of common semiconductors}



\author{David Porlles}
\affiliation{Department of Physics, PUC-Rio, 22451-900 Rio de Janeiro, Brazil}

\author{Wei Chen}
\affiliation{Department of Physics, PUC-Rio, 22451-900 Rio de Janeiro, Brazil}

\date{\rm\today}

\begin{abstract}

The quantum geometric properties of typical diamond-type (C, Si, Ge) and zincblende-type (GaAs, InP, etc) semiconductors are investigated by means of the $sp^{3}s^{\ast}$ tight-binding model, which allows to calculate the quantum metric of the valence band states throughout the entire Brillouin zone. The global maximum of the metric is at the $\Gamma$ point, but other differential geometric properties like Ricci scalar, Ricci tensor, and Einstein tensor are found to vary significantly in the momentum space, indicating a highly distorted momentum space manifold. The momentum integration of the quantum metric further yields the gauge-invariant part of the spread of valence band Wannier function, whose value agrees well with that experimentally extracted from an optical sum rule of the dielectric function. Furthermore, the dependence of these geometric properties on the energy gap offers a way to quantify the quantum criticality of these common semiconductors. 

\end{abstract}

\maketitle


The notion of quantum metric emerges recently as the mechanism behind many physical properties of semiconductors and insulators, which has stimulated a great deal of interest in this quantity. This concept of quantum metric is introduced by considering two filled valence band states at slightly different momenta, whose product expanded over the small momentum difference yields the quantum metric\cite{Provost80}. Having this metric defined, one may further introduce differential geometric quantities by treating the momentum space as a Euclidean manifold, which has revealed several interesting differential geometric properties, such as the constant Ricci scalar and cosmological constant in topological materials\cite{Matsuura10,Chen24_generic_TI_TSC}. Furthermore, because this metric enters the optical transition matrix elements of interband transitions\cite{Ozawa18}, it manifests in many optical responses like optical conductivity\cite{Ahn22}, capacitance\cite{Komissarov24}, dielectric function, charge susceptibility, refractive index, absorption coefficient, reflectance, and transmittance\cite{Chen25_optical_marker}. This implies that the quantum metric determines a large part of dielectric and optical properties that can even be perceived by human eyes in the macroscopic scale.

Another significance of the quantum metric is that the momentum integration of the quantum metric is equivalently the gauge-invariant part of the spread of Wannier function summing over valence band states, which characterizes how insulating is the material from a real space perspective\cite{Marzari97,Marzari12,Souza08}. This spread can be generalized to disordered systems by defining it locally on each lattice site, rendering a localization marker\cite{Marrazzo17_2} or fidelity marker\cite{deSousa23_fidelity_marker}. By combing with the aforementioned dielectric and optical responses, it is proposed recently that the dielectric function in three dimensions and the absorbance in two dimensions can be used to measure the spread even under the influence of disorder\cite{Cardenas24_spread_Wannier,Cardenas25_absorbance_marker}, bringing the experimental detection of the spread to reality.

However, despite these important features, the quantum metric of the most common semiconductors have not been investigated in detail. Given the vast applications in information technology, the quantum geometry in the typical diamond-type semiconductors of group IV like C, Si, and Ge, and the zincblende-type semiconductors of group III-V like GaAs and InP certainly deserves to be investigated. In this paper, we investigate the quantum geometric properties of these common semiconductors by means of the $sp^{3}s^{\ast}$ tight-binding model\cite{Vogl83}. This $sp^{3}s^{\ast}$ model is known to give a reasonable description of the band structure of both the valence and conduction bands by using very few band parameters, and has been applied to investigate numerous properties like nanostructures\cite{Lippens89,GuzmanVerri07}, strain effects\cite{Sun07}, and superlattice\cite{Smith90}, among many others. In addition, the model has also been used to calculate the Berry curvature of silicon, revealing a network of Berry flux lines that connect high symmetry points\cite{Shtyk20}. Our goal is to investigate the quantum metric of this model throughout the entire Brillouin zone (BZ), quantify the momentum profile of the metric and other differential geometric properties, as well as calculating the dielectric function and extract the spread of Wannier function. Furthermore, investigating the dependence of these geometric properties on the energy gap helps to quantify the quantum criticality of these materials, which reveals a rather peculiar behavior of the metric and the spread of Wannier function when these semiconductors are driven towards the critical point.



We consider fully gapped semiconductors with direct or indirect band gaps, and use $|n\rangle$ to represent the $N_{-}$ valence band states with energy $\varepsilon_{n}<0$, $|m\rangle$ to represent the $N_{+}$ conduction band states with energy $\varepsilon_{m}>0$, and $|\ell\rangle$ to enumerate both states. The Bloch state under consideration is the fully antisymmetric many-body valence band state given by
\begin{eqnarray}
|u^{\rm val}({\bf k})\rangle=\frac{1}{\sqrt{N_{-}!}}\epsilon^{n_{1}n_{2}...n_{N-}}|n_{1}\rangle|n_{2}\rangle...|n_{N_{-}}\rangle.\;\;\;
\label{psi_val}
\end{eqnarray}
The quantum metric is defined from the overlap of two such states at neighboring momenta\cite{Provost80} 
\begin{eqnarray}
|\langle u^{\rm val}({\bf k})|u^{\rm val}({\bf k+\delta k})\rangle|=1-\frac{1}{2}g_{\mu\nu}({\bf k})\delta k^{\mu}\delta k^{\nu},
\label{uval_gmunu}
\end{eqnarray}
which can be expressed by\cite{Matsuura10,vonGersdorff21_metric_curvature} 
\begin{eqnarray}
&&g_{\mu\nu}({\bf k})=\frac{1}{2}\langle \partial_{\mu}u^{\rm val}|\partial_{\nu}u^{\rm val}\rangle+\frac{1}{2}\langle \partial_{\nu}u^{\rm val}|\partial_{\mu}u^{\rm val}\rangle
\nonumber \\
&&-\langle \partial_{\mu}u^{\rm val}|u^{\rm val}\rangle \langle u^{\rm val}|\partial_{\nu}u^{\rm val}\rangle
\nonumber \\
&&=
\frac{1}{2}\sum_{nm}\left[\langle \partial_{\mu}n|m\rangle\langle m|\partial_{\nu}n\rangle+\langle \partial_{\nu}n|m\rangle\langle m|\partial_{\mu}n\rangle\right]
\nonumber \\
&&=\frac{1}{2}\sum_{nm}\left[\frac{\langle n|\partial_{\mu}H|m\rangle\langle m|\partial_{\nu}H|n\rangle+\langle n|\partial_{\nu}H|m\rangle\langle m|\partial_{\mu}H|n\rangle}{(\varepsilon_{n}-\varepsilon_{m})^{2}}\right]
\nonumber \\
&&=\sum_{nm}g_{\mu\mu}^{nm},
\label{gmunu_T0}
\end{eqnarray}
with $\partial_{\mu}\equiv\partial/\partial k^{\mu}$. Furthermore, 
momentum integral of the quantum metric\cite{Souza08,Marzari97,Marzari12} 
\begin{eqnarray}
&&{\cal G}_{\mu\nu}=\int\frac{d^{3}{\bf k}}{(2\pi)^{3}}g_{\mu\nu}({\bf k}),
\label{Gmunu_definition}
\end{eqnarray}
of which we call the fidelity number\cite{deSousa23_fidelity_marker}, will be particularly important for the optical sum rule described below.

Once the metric calculated, we further investigate the differential geometrical quantities that describe the Euclidean manifold of momentum space, including the Christoffel symbol $\Gamma_{\mu\nu}^{\lambda}$, Riemann tensor $R^{\rho}_{\;\sigma\mu\nu}$, Ricci tensor $R_{\mu\nu}$, Ricci scalar $R$, and Einstein tensor $G_{\mu\nu}$ calculated by\cite{Carroll03}
\begin{eqnarray}
&&\Gamma_{\mu\nu}^{\lambda}=\frac{1}{2}g^{\lambda\sigma}(\partial_{\mu}g_{\nu\sigma}
+\partial_{\nu}g_{\sigma\mu}-\partial_{\sigma}g_{\mu\nu}),
\nonumber \\
&&R^{\rho}_{\;\sigma\mu\nu}=\partial_{\mu}\Gamma_{\nu\sigma}^{\rho}-\partial_{\nu}\Gamma_{\mu\sigma}^{\rho}
+\Gamma_{\mu\lambda}^{\rho}\Gamma_{\nu\sigma}^{\lambda}-\Gamma_{\nu\lambda}^{\rho}\Gamma_{\mu\sigma}^{\lambda},
\nonumber \\
&&R_{\mu\nu}=R^{\lambda}_{\;\mu\lambda\nu},\;\;\;
R=g^{\mu\nu}R_{\nu\mu},
\nonumber \\
&&G_{\mu\nu}=R_{\mu\nu}-\frac{1}{2}Rg_{\mu\nu},
\label{Christoffel_Riemann_Ricci}
\end{eqnarray} 
where $g^{\nu\rho}$ satisfies $g_{\mu\nu}g^{\nu\rho}=\delta_{\mu}^{\rho}$. Note that the metric itself can be measured by pump-probe experiments via detecting the loss of particle density in the valence bands at momentum ${\bf k}$ caused by the pump field\cite{vonGersdorff21_metric_curvature}, which may be detected via time-resolved and angle-resolved photoemission spectroscopy (tr-ARPES)\cite{Gierz13}. Once momentum profile of $g_{\mu\nu}$ is measured, other quantities in Eq.~(\ref{Christoffel_Riemann_Ricci}) can be obtained through straightforward momentum derivative.


For either the valence band $\ell=n$ or conduction band $\ell=m$, the periodic part of the Bloch wave function $\langle{\bf r}|\ell\rangle=\ell({\bf r})=e^{-i{\bf k\cdot r}/\hbar}\psi_{\ell}^{\bf k}({\bf r})$ satisfies $\ell({\bf r})=\ell({\bf r+R})$, with ${\bf r}$ and ${\bf R}$ the position and Bravais lattice vectors, respectively. From this periodicity, one introduces the Wannier state $|{\bf R}\ell\rangle$ by
\begin{eqnarray}
&&|{\bf R} \ell\rangle=\sum_{\bf k}e^{i {\bf k}\cdot({\hat{\bf r}}-{\bf R})/\hbar}|\ell\rangle.
\label{Wannier_basis}
\end{eqnarray}
A characteristic quantity that measures the range of Wannier function is the second cumulant of the charge distribution, called the spread of Wannier function\cite{Marzari97,Marzari12,Souza08}. We are particularly interested in the gauge-invariant part of the spread defined from summing over all the valence band states given by the trace of the fidelity number in Eq.~(\ref{Gmunu_definition})
\begin{eqnarray}
&&\Omega_{I}=\sum_{n}\left[\langle{\bf 0}n|r^{2}|{\bf 0}n\rangle-\sum_{{\bf R}n'}|\langle{\bf R}n'|{\bf r}|{\bf 0}n\rangle|^{2}\right]
\nonumber \\
&&=\frac{V_{\rm cell}}{\hbar}\,{\rm Tr}\,{\cal G}_{\mu\nu}=\frac{V_{\rm cell}}{\hbar}\sum_{\mu}{\cal G}_{\mu\mu}\,.
\label{OmegaI_trace_Gmunu}
\end{eqnarray}
The key to the measurement of $\Omega_{I}$ and ${\rm Tr}\,{\cal G}_{\mu\nu}$ is the dielectric function: Given the relation between the complex dielectric function and complex optical conductivity
\begin{eqnarray}
\varepsilon(\epsilon)=1+i\frac{\sigma(\omega)}{\varepsilon_{0}\omega}=\varepsilon_{1}+i\varepsilon_{2}.
\end{eqnarray}
Because the elements of the quantum metric $g_{\mu\mu}^{nm}$ in Eq.~(\ref{gmunu_T0}) are the optical transition matrix elements of the real part of optical conductivity $\sigma_{1}(\omega)$, one sees that\cite{Graf95,Chen25_optical_marker} 
\begin{eqnarray}
&&\varepsilon_{2}(\omega)=\frac{\sigma_{1}(\omega)}{\varepsilon_{0}\omega}
\nonumber \\
&&=\frac{\pi e^{2}}{\varepsilon_{0}\hbar^{2}}\int\frac{d^{3}{\bf k}}{(2\pi)^{3}}\sum_{nm}g_{\mu\mu}^{nm}\delta\left(\omega+\frac{\varepsilon_{n}}{\hbar}-\frac{\varepsilon_{m}}{\hbar}\right).
\label{epsilon2_gmumu_int}
\end{eqnarray}
As a result, $\Omega_{I}$ and ${\rm Tr}\,{\cal G}_{\mu\nu}$ are directly determined by the frequency integrals of the imaginary part of dielectric function $\varepsilon_{2}(\omega)$
\begin{eqnarray}
&&\int_{0}^{\infty}d(\hbar\omega)\,\varepsilon_{2}(\omega)
\equiv\int_{0}^{\infty}d({\rm eV})\,\varepsilon_{2}({\rm eV})
=\nu\times {\rm eV},
\nonumber \\
&&\Omega_{I}=\lim_{T\rightarrow 0}\frac{V_{\rm cell}\varepsilon_{0}}{\pi e^{2}}\,{\rm eV}\times 3\nu
\nonumber \\
&&\;\;\;\;\;=\lim_{T\rightarrow 0}\frac{V_{\rm cell}}{\AA}\times 1.7591\times 10^{-3}\times 3\nu,
\nonumber \\
&&{\rm Tr}{\cal G}_{\mu\nu}=\lim_{T\rightarrow 0}\frac{\hbar}{\AA}\times 1.7591\times 10^{-3}\times 3\nu,
\label{spread_3D_general_formula}
\end{eqnarray}
which is readily applicable to common semiconductors\cite{Cardenas24_spread_Wannier}. 



To quantify the quantum geometric properties for diamond-type and zincblende-type semiconductors, we choose to adopt the $sp^{3}s^{\ast}$ model of Vogl et al that is relatively simple, and gives a fairly accurate description on both the valence bands and conduction bands that are relevant to the optical absorption\cite{Vogl83,Graf95}. The $10\times 10$ Hamiltonian of the $sp^{3}s^{\ast}$ model in the basis denoted by $\left\{sa,sc,p_{x}a,p_{y}a,p_{z}a,p_{x}c,p_{y}c,p_{z}c,s^{\ast}a,s^{\ast}c\right\}$, is given by
\begin{widetext}
\begin{eqnarray}
&&H({\bf k})=
\nonumber \\
&&\left(\begin{array}{cccccccccc}
E_{sa} & V_{s,s}g_{0} & 0 & 0 & 0 & V_{sa,pc}g_{1} & V_{sa,pc}g_{2} & V_{sa,pc}g_{3} & 0 & 0 \\
V_{s,s}g_{0}^{\ast} & E_{sc} & -V_{pa,sc}g_{1}^{\ast} & -V_{pa,sc}g_{2}^{\ast} & -V_{pa,sc}g_{3}^{\ast} & 0 & 0 & 0 & 0 & 0 \\
0 & -V_{pa,sc}g_{1} & E_{pa} & 0 & 0 & V_{x,x}g_{0} & V_{x,y}g_{3} & V_{x,y}g_{2} & 0 & -V_{pa,s^{\ast}c}g_{1} \\
0 & -V_{pa,sc}g_{2} & 0 & E_{pa} & 0 & V_{x,y}g_{3} & V_{x,x}g_{0} & V_{x,y}g_{1} & 0 & -V_{pa,s^{\ast}c}g_{2} \\
0 & -V_{pa,sc}g_{3} & 0 & 0 & E_{pa} & V_{x,y}g_{2} & V_{x,y}g_{1} & V_{x,x}g_{0} & 0 & -V_{pa,s^{\ast}c}g_{3} \\
V_{sa,pc}g_{1}^{\ast} & 0 & V_{x,x}g_{0}^{\ast} & V_{x,y}g_{3}^{\ast} & V_{x,y}g_{2}^{\ast} & E_{pc} & 0 & 0 & V_{s^{\ast}a,pc}g_{1}^{\ast} & 0 \\
V_{sa,pc}g_{2}^{\ast} & 0 & V_{x,y}g_{3}^{\ast} & V_{x,x}g_{0}^{\ast} & V_{x,y}g_{1}^{\ast} & 0 & E_{pc} & 0 & V_{s^{\ast}a,pc}g_{2}^{\ast} & 0 \\
V_{sa,pc}g_{3}^{\ast} & 0 & V_{x,y}g_{2}^{\ast} & V_{x,y}g_{1}^{\ast} & V_{x,x}g_{0}^{\ast} & 0 & 0 & E_{pc} & V_{s^{\ast}a,pc}g_{3}^{\ast} & 0 \\
0 & 0 & 0 & 0 & 0 & V_{s^{\ast}a,pc}g_{1} & V_{s^{\ast}a,pc}g_{2} & V_{s^{\ast}a,pc}g_{3} & E_{s^{\ast}a} & 0 \\
0 & 0 & -V_{pa,s^{\ast}c}g_{1}^{\ast} & -V_{pa,s^{\ast}c}g_{2}^{\ast} & -V_{pa,s^{\ast}c}g_{3}^{\ast} & 0 & 0 & 0 & 0 & E_{s^{\ast}c}
\end{array}\right),
\nonumber \\
&&g_{0}({\bf k})=\;\;\;\cos\frac{k_{x}}{4}\cos\frac{k_{y}}{4}\cos\frac{k_{z}}{4}-i\sin\frac{k_{x}}{4}\sin\frac{k_{y}}{4}\sin\frac{k_{z}}{4},
\;\;\;\;\;g_{1}({\bf k})=-\cos\frac{k_{x}}{4}\sin\frac{k_{y}}{4}\sin\frac{k_{z}}{4}+i\sin\frac{k_{x}}{4}\cos\frac{k_{y}}{4}\cos\frac{k_{z}}{4},
\nonumber \\
&&g_{2}({\bf k})=-\sin\frac{k_{x}}{4}\cos\frac{k_{y}}{4}\sin\frac{k_{z}}{4}+i\cos\frac{k_{x}}{4}\sin\frac{k_{y}}{4}\cos\frac{k_{z}}{4},
\;\;\;\;\;g_{3}({\bf k})=-\sin\frac{k_{x}}{4}\sin\frac{k_{y}}{4}\cos\frac{k_{z}}{4}+i\cos\frac{k_{x}}{4}\cos\frac{k_{y}}{4}\sin\frac{k_{z}}{4}.
\end{eqnarray}
\end{widetext}
The values of the band parameters for all the semiconductors under investigation are tabulated in the original work\cite{Vogl83}, and the derivative of the Hamiltonian over momentum $\partial_{\mu}H({\bf k})$ is straightforward, which can then be inserted into Eq.~(\ref{gmunu_T0}) to calculate the quantum metric $g_{\mu\nu}({\bf k})$ on evenly spaced mesh points. The derivatives $\partial_{\mu}g_{\lambda\nu}$ and $\partial_{\mu}\Gamma_{\nu\lambda}^{\rho}$ in Eq.~(\ref{Christoffel_Riemann_Ricci}) can then be evaluated by means of central difference 
\begin{eqnarray}
\partial_{\mu}A({\bf k})=\frac{A({\bf k}+\Delta k{\hat{\boldsymbol\mu}})-A({\bf k}-\Delta k{\hat{\boldsymbol\mu}})}{2\Delta k},
\end{eqnarray}
with $A=\left\{g_{\lambda\nu},\Gamma_{\nu\lambda}^{\rho}\right\}$. We use $32\times 32\times 32$ mesh points that are sufficient to obtain smooth profiles of the geometric quantities in Eq.~(\ref{Christoffel_Riemann_Ricci}).


\begin{figure*}[ht]
\begin{center}
\includegraphics[clip=true,width=1.99\columnwidth]{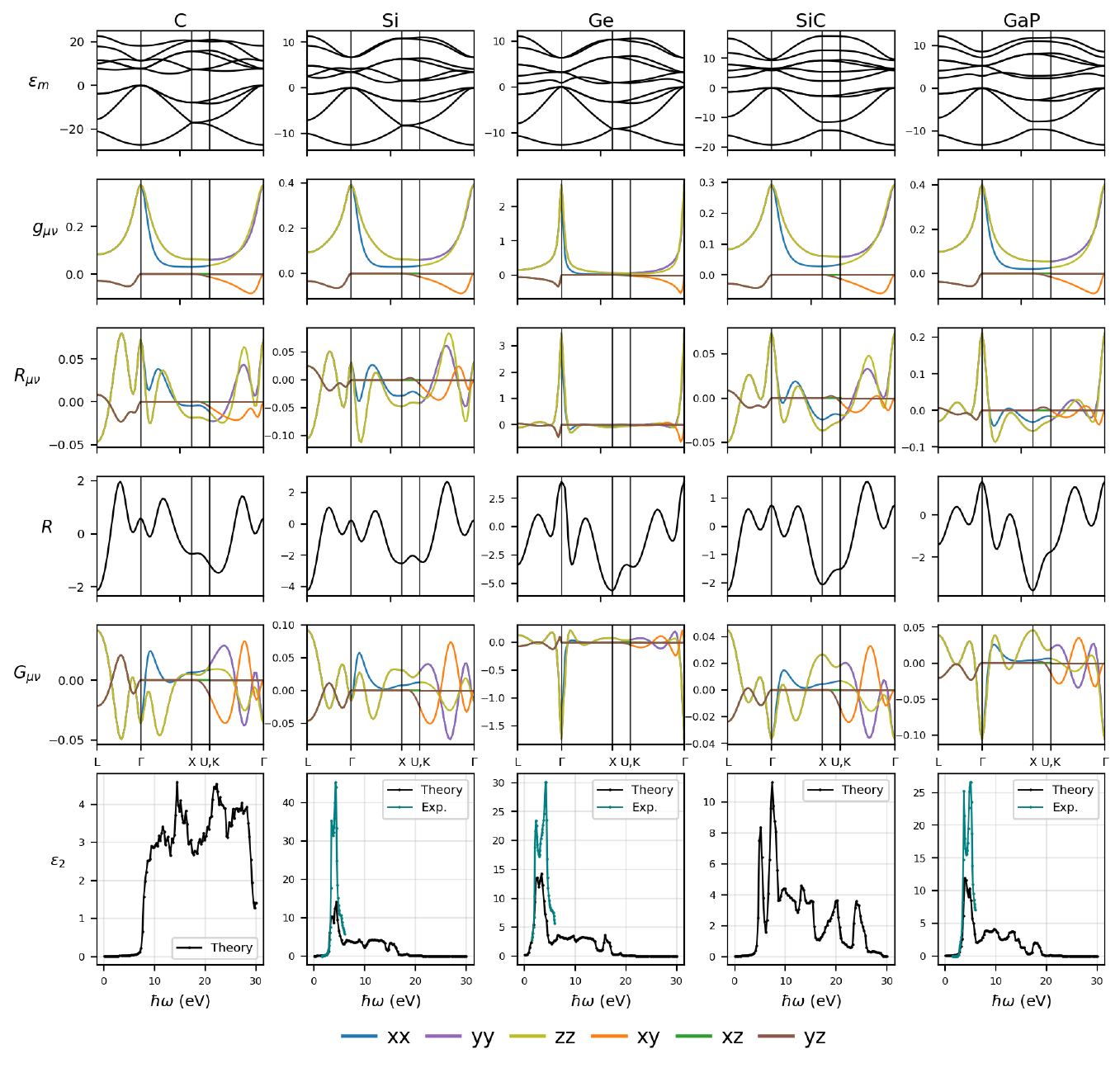}
\caption{Numerical results for the quantum geometric properties of the 5 indirect band gap semiconductors C, Si, Ge, SiC, and Gap obtained from the $sp^{3}s^{\ast}$ tight-binding model, plotted along the high symmetry line $L-\Gamma-X-(U,K)-\Gamma$. From top to bottom, we present the band structure $(\varepsilon_{n},\varepsilon_{m})$, quantum metric $g_{\mu\nu}$, Ricci tensor $R_{\mu\nu}$, Ricci scalar $R$, and Einstein tensor $G_{\mu\nu}$, with $\mu\nu=\left\{xx,xy,xz,yy,yz,zz\right\}$ indicated by different colors. The last row shows that imaginary part of dielectric function $\varepsilon_{2}(\omega)$, together with the experimental data available for some of these materials listed in Ref.~\onlinecite{Aspnes83} shown as dashed lines.
} 
\label{fig:indirect_gap_SM_figure}
\end{center}
\end{figure*}

\begin{figure*}[ht]
\begin{center}
\includegraphics[clip=true,width=1.9\columnwidth]{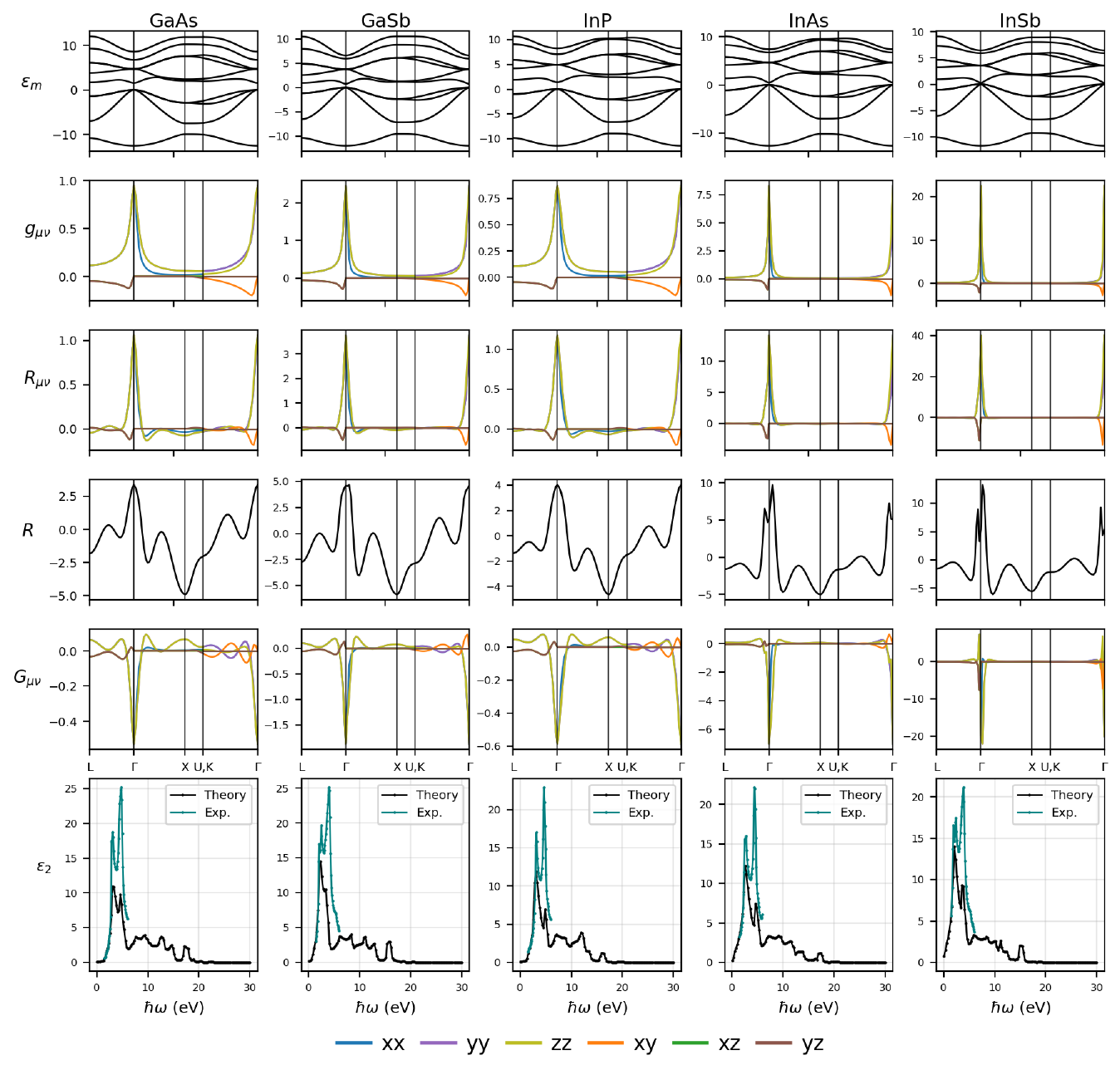}
\caption{Same as Fig.~\ref{fig:indirect_gap_SM_figure}, but for the 5 direct band gap semiconductors GaAs, GaSb, InP, InAs, and InSb.} 
\label{fig:direct_gap_SM_figure}
\end{center}
\end{figure*}


For concreteness, we investigate the geometric properties along the high symmetry line $L-\Gamma-X-(U,K)-\Gamma$, with the five high symmetry points located at $L=\left(1/2,1/2,1/2\right)$, $\Gamma=(0,0,0)$, $X=(1,0,0)$, $U=(1,1/4,1/4)$, and $K=(3/4,3/4,0)$. The results for the indirect band gap semiconductors C, Si, Ge, SiC, and GaP are presented in Fig.~\ref{fig:indirect_gap_SM_figure}, and for the direct band gap semiconductors GaAs, GaSb, InP, In As, and InSb are presented in Fig.~\ref{fig:direct_gap_SM_figure}. We present the geometric properties that are of rank two, including the quantum metric $g_{\mu\nu}$, Ricci tensor $R_{\mu\nu}$, and Einstein tensor $G_{\mu\nu}$, all of which are symmetric under the exchange of the two indices $\mu\leftrightarrow\nu$, so there are only six independent elements $\mu\nu=\left\{xx,xy,xz,yy,yz,zz\right\}$. We also present the band structure of the four valence bands $\varepsilon_{n}$ and six conduction bands $\varepsilon_{m}$, and the Ricci scalar $R$. Finally, we also plot the imaginary part of dielectric function $\varepsilon_{2}(\omega)$ and the experimental data for some of these materials tabulated in Ref.~\onlinecite{Aspnes83}.



For all these 10 materials, the global maximum of the diagonal elements of the quantum metric $g_{\mu\mu}$ is at the $\Gamma$ point, indicating that the valence band state $|u^{\rm val}({\bf k})\rangle$ changes most dramatically with momentum at the $\Gamma$ point. Moreover, the magnitude of $g_{\mu\mu}$ at $\Gamma$ is found to be much larger in InSb that has the smallest band gap $\Delta_{\Gamma}$ at the $\Gamma$ point. The width of the $g_{\mu\mu}$ peak at $\Gamma$ is also found to be much smaller in InSb, a phenomena similar to the narrowing and divergence of Berry curvature or spin Berry curvature in topological materials\cite{Chen17,Chen19_universality}, in which the divergence of quantum metric also occurs due to a metric-curvature correspondence\cite{vonGersdorff21_metric_curvature}.



The Ricci tensor $R_{\mu\nu}$ and scalar $R$ have profiles that fluctuate dramatically in the whole BZ for all the 10 semiconductors, and none of the high symmetry points seems to be particularly important for these quantities. In particular, the Ricci scalar $R$ has positive and negative signs in different regions of the BZ, meaning that the manifold in curved in opposite manners in different regions, and hence a highly distorted momentum space manifold. This is very different from Dirac models of topological materials that has only positive Ricci scalar throughout the entire BZ\cite{Matsuura10,Chen24_generic_TI_TSC}. In addition, the Ricci scalar at the $\Gamma$ point $R(\Gamma)$ also increases as the energy gap $\Delta_{\Gamma}$ reduces, suggesting that it also senses the quantum criticality near the critical point. Finally, the Einstein tensor $G_{\mu\nu}=R_{\mu\nu}-Rg_{\mu\nu}/2$ and its two components $R_{\mu\nu}$ and $Rg_{\mu\nu}/2$ are all nonzero and of the same order of magnitude, indicating that most part of the manifold does not satisfy the vacuum Einstein equation $G_{\mu\nu}\neq 0$, unlike the topological materials\cite{Chen24_generic_TI_TSC}.

To compare with the experimental dara\cite{Aspnes83}, the $\varepsilon_{2}(\omega)$ calculated by Eq.~(\ref{epsilon2_gmumu_int}) is multiplied by a factor of 8 to account for the 8 atoms per unit cell. We find that the resulting theoretical $\varepsilon_{2}(\omega)$ has a similar frequency dependence as the experimental one, but the overall magnitude is smaller. This feature is known in all the theoretical calculations of $\varepsilon_{2}(\omega)$ based on interband transitions, which does not capture the extra spectral weight caused by excitons in the experimental data\cite{Graf95,Yu10}. Nevertheless, our result indicates that the $sp^{3}s^{\ast}$ model does capture the interband transition part reasonably well, suggesting the adequacy of this model. 


In Table \ref{tab:sp3s_model_data}, we summarize the volume of the unit cell $V_{\rm cell}$, the frequency integration of the dielectric function $\nu$, trace of the fidelity number ${\rm Tr}{\cal G}_{\mu\nu}$, the spread $\Omega_{I}$, the energy gap at the $\Gamma$ point $\Delta_{\Gamma}$ (not the overall band gap for indirect band gap semiconductors), and the dimensionless ration $\Omega_{I}^{3/2}/V_{\rm cell}$ for all the 10 semiconductors investigated. Compared to the values of ${\rm Tr}\,{\cal G}_{\mu\nu}$ and $\Omega_{I}$ extracted from the experimental data\cite{Cardenas24_spread_Wannier}, the $sp^{3}s^{\ast}$ model gives about $70\%$ to $80\%$ of the experimental value. In fact, the theoretical values of ${\rm Tr}\,{\cal G}_{\mu\nu}$ and $\Omega_{I}$ may be closer to the true values in real semiconductors, since the experimental data likely overestimates the ${\rm Tr}\,{\cal G}_{\mu\nu}$ and $\Omega_{I}$ due to the extra spectral weight in $\varepsilon_{2}(\omega)$ caused by excitons as mentioned earlier.


\begin{table}[ht]
  \begin{center}
    \caption{Volume of the unit cell $V_{\rm cell}$, frequency-integration $\nu$ of the imaginary part of dielectric function, trace of the fidelity number ${\rm Tr}{\cal G}_{\mu\nu}$, spread of the valence band Wannier functions $\Omega_{I}$, and the gap $\Delta_{\Gamma}$ at the $\Gamma$ point obtained from the $sp^{3}s^{\ast}$ model for the 10 materials under investigation.}
    \label{tab:sp3s_model_data}
    \begin{tabular}{ c c c c c c c}
    \hline
    {\rm Mat}. & $V_{\rm cell}(\AA^{3})$ & $\nu$ & ${\rm Tr}{\cal G}_{\mu\nu}(\hbar/\AA)$ & $\Omega_{I}(\AA^{2})$ & $\Delta_{\Gamma}(eV)$ & $\Omega_{I}^{3/2}/V_{\rm cell}$ \\ \hline
    C & 45.5 & 92.3 & 0.487 & 22.2 & 7.68 & 2.30 \\ 
    Si & 160.1 & 62.9 & 0.332 & 53.2 & 3.43 & 2.42  \\ 
    Ge & 181.3 & 67.6 & 0.357 & 64.7 & 3.22 & 2.87 \\ 
    SiC & 82.9 & 73.0 & 0.385 & 31.9 & 5.90 & 2.17 \\ 
    GaP & 161.9 & 55.5 & 0.293 & 47.4 & 2.88 & 2.02 \\ 
    GaAs & 180.4 & 57.2 & 0.302 & 54.5 & 1.55 & 2.23 \\ 
    GaSb & 226.5 & 57.6 & 0.304 & 68.9 & 0.78 & 2.53 \\ 
    InP & 202.3 & 49.5 & 0.261 & 52.8 & 1.41 & 1.90 \\ 
    InAs & 222.5 & 54.2 & 0.286 & 63.6 & 0.43 & 2.28 \\ 
    InSb & 272.1 & 55.0 & 0.290 & 78.9 & 0.23 & 2.58 \\ 
    \hline
  \end{tabular}
  \end{center}
\end{table}



\begin{figure}[ht]
\begin{center}
\includegraphics[clip=true,width=0.99\columnwidth]{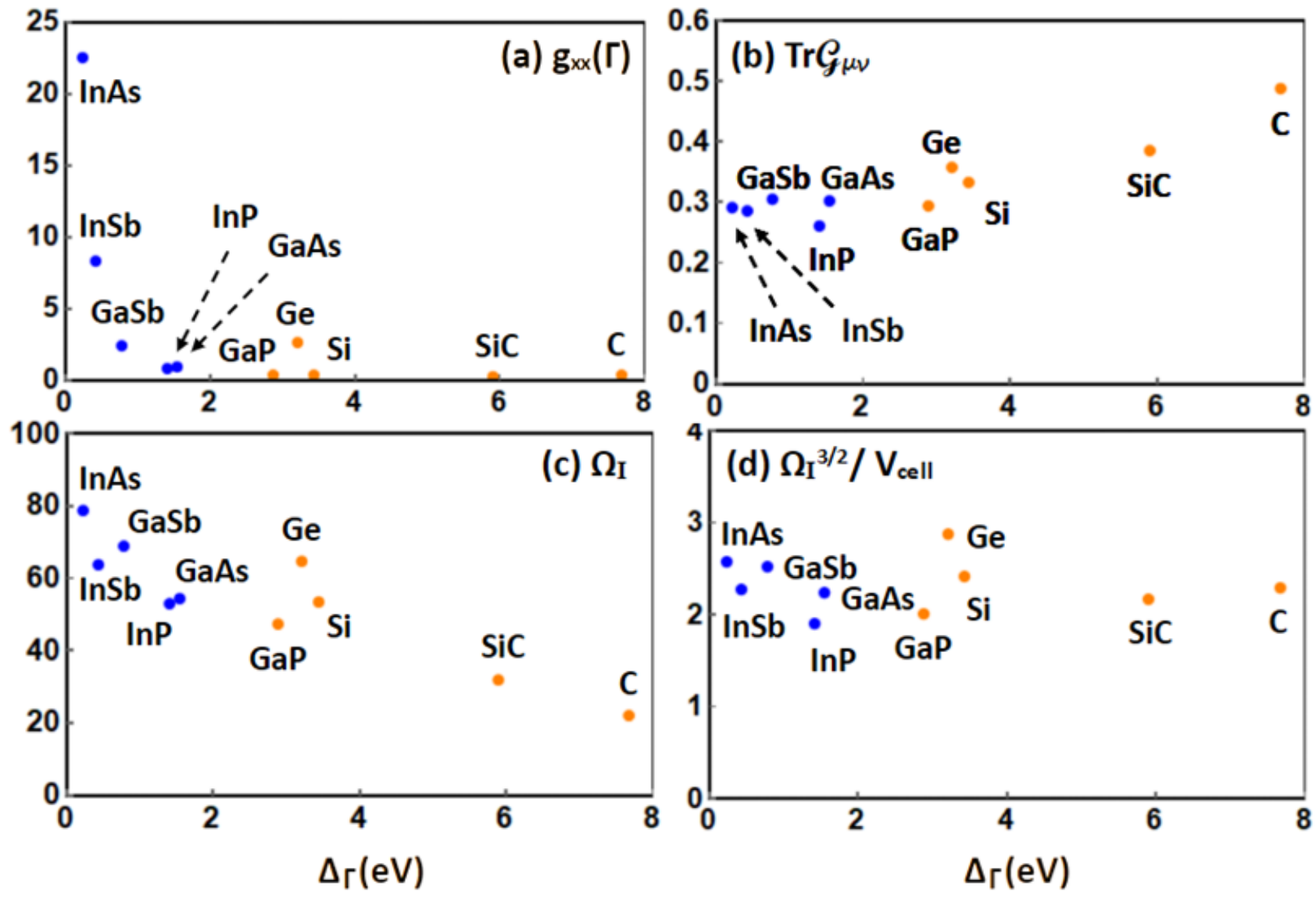}
\caption{(a) The quantum metric at the $\Gamma$ point $g_{xx}(\Gamma)$, (b) trace of fidelity number ${\rm Tr}{\cal G}_{\mu\nu}$, (c) the gauge-invariant part of the spread of valence band Wannier function $\Omega_{I}$, and (d) the dimensionless ratio $\Omega_{I}^{3/2}/V_{\rm cell}$ versus the gap at the $\Gamma$ point $\Delta_{\Gamma}$ for the ten semiconductors under investigation. } 
\label{fig:TrG_OmegaI_vs_gap}
\end{center}
\end{figure}


In Fig.~\ref{fig:TrG_OmegaI_vs_gap}, we present the quantum geometric properties as a function of the energy gap $\Delta_{\Gamma}$ at the $\Gamma$ point for the 10 semiconductors under investigation. We choose to present these properties as a function of $\Delta_{\Gamma}$ because it is the energy difference that determines the maximal $g_{\mu\mu}$ at the $\Gamma$ point, and hence characterizing how close are the semiconductors to the critical point, even though $\Delta_{\Gamma}$ is not the global band gap for the 5 indirect band gap semiconductors. This strategy allows us to describe the quantum criticality of these semiconductors in a unified manner, thereby quantifies the critical behavior of the $sp^{3}s^{\ast}$ model. Indeed, from Fig.~\ref{fig:TrG_OmegaI_vs_gap} (a), one sees that the metric $g_{xx}(\Gamma)$ at the $\Gamma$ point increases as $\Delta_{\Gamma}$ decreases, manifesting a tendency of divergence as $\Delta_{\Gamma}\rightarrow 0$. This behavior is in line with the physical picture that the quantum metric can also be viewed as a fidelity susceptibility for the valence band state $|u^{\rm val}({\bf k})\rangle$ treating momentum ${\bf k}$ as a tuning parameter\cite{You07,Zanardi07,Gu08}, which diverges at the critical point where the energy gap $\Delta_{\Gamma}$ closes. On the other hand, the momentum integration of the quantum metric ${\rm Tr}{\cal G}_{\mu\nu}$ shown in Fig.~\ref{fig:TrG_OmegaI_vs_gap} (b) saturates to about 0.3 as the critical point $\Delta_{\Gamma}\rightarrow 0$, and increases with $\Delta_{\Gamma}$. This implies that the quantum metric in the rest of the BZ is not much affected by whether the $\Gamma$ point is close to the critical point or not, and still integrates to a finite number, similar to what occurs in topological materials\cite{Chen17,Chen19_universality,Molignini19}.

Turning to the spread of Wannier function $\Omega_{I}$ in Fig.~\ref{fig:TrG_OmegaI_vs_gap} (c), note that according to Eq.~(\ref{OmegaI_trace_Gmunu}), the dependence of $\Omega_{I}$ on $\Delta_{\Gamma}$ is determined by how the ${\rm Tr}{\cal G}_{\mu\nu}$ and $V_{\rm cell}$ depend on $\Delta_{\Gamma}$. The combined effect of the two yields an $\Omega_{I}$ that decreases with $\Delta_{\Gamma}$, which seems to be in accordance with the intuitive picture that because a smaller gap material is less insulating, its Wannier function should be more extended and have more overlap on the neighboring sites\cite{Marrazzo17_2}. However, we find that this intuitive picture is not entirely correct if one takes the volume of the unit cell $V_{\rm cell}$ into account. To quantify the spread $\Omega_{I}$ relative to the $V_{\rm cell}$, in Fig.~\ref{fig:TrG_OmegaI_vs_gap} (d) we present the dimensionless ratio $\Omega_{I}^{3/2}/V_{\rm cell}$, which remains roughly constant as a function of $\Delta_{\Gamma}$, taking the value between 2 and 3. This remarkable result indicates that as a semiconductor approaches the critical point, the overlap between valence band Wannier functions on neighboring unit cells remains roughly constant. Thus the spread $\Omega_{I}$ and the ratio $\Omega_{I}^{3/2}/V_{\rm cell}$ themselves are not sufficient to judge how close is the material to the critical point, but rather it is the Fourier transform of the quantum metric that inherits the divergence of the metric at the $\Gamma$ point that can capture the transition\cite{deSousa23_fidelity_marker}.


In summary, we use the $sp^{3}s^{\ast}$ tight-binding model to investigate the quantum geometry of common diamond-type and zincblende-type semiconductors. The diagonal elements of quantum metric $g_{\mu\mu}$ are found to be maximal at the $\Gamma$ point, and the wildly fluctuating Ricci scalar indicates that the momentum space manifold is highly distorted, yielding vacuum Einstein equation not satisfied. The spread of valence band Wannier function extracted theoretically agrees qualitatively well with that obtained from the experimental data of dielectric function. Through investigating the dependence of these quantities on the energy gap at the $\Gamma$ point, we reveal that the overlap between Wannier functions on neighboring sites remains roughly constant as the semiconductor approaches the critical point. Besides, the quantum metric integrated over the BZ actually increases as the semiconductor moves away from the critical point. We anticipate that the methodology presented in our work, which introduces the differential geometric properties into semiconductors in an experimentally observable manner, and moreover quantifies the critical behavior of semiconductors, can be applied to investigate a wide range of other insulating materials. 


\begin{acknowledgments}

We acknowledge the support of the INCT project Advanced Quantum Materials, 
involving the Brazilian agencies CNPq (Proc. 408766/2024-7), FAPESP, and CAPES. W.C. acknowledges the financial support of the productivity in research fellowship from CNPq. 

\end{acknowledgments}

\section*{Data availability}
The data that support the findings of this study are available from the corresponding authors upon reasonable request.

\bibliography{Literatur}

\end{document}